# Record statistics for a discrete-time random walk with correlated steps

Michael J Kearney

*Senate House, University of Surrey, Guildford, Surrey, GU2 7XH, UK*

*Abstract*

The characterization of record events is considered for a discrete-time random walk model with long-term memory arising from correlations between successive steps. An important feature is that the correlations are strong enough to give rise to super-diffusivity and transience. Various quantities related to record statistics are calculated exactly, highlighting important differences in behaviour from the simple random walk with independent steps.

## 1. Introduction

A great deal is known about the statistics of record events for discrete-time processes where the variables are independent and identically distributed; see e.g. [1]. More recently, attention has shifted to consider records in the context of correlated time series such as random walks but still with independent steps (increments). For symmetric walks, or walks with constant drift, much progress has been made concerning the number of records in a given observation interval, as well as details of their duration [2-7]. Such results find many areas of application in the physical, life and even social sciences; see e.g. [8] and references therein.

A natural extension, and the theme of this paper, is to consider records in a discrete-time random walk model with *strongly correlated steps* resulting in positive reinforcement and persistence of motion, to the extent that the overall dynamics features super-diffusivity and transience. This will impact significantly on the record statistics as compared to random walks with independent steps. To define the model, let us denote the position of a walker by the integer variable $\Delta_S$, where $S \geq 0$ is the number of steps taken. Starting at $\Delta_0 = 0$, the evolution follows,

$$\Delta_{S+1} = \Delta_S + \sigma_{S+1}; \qquad \Delta_S = \sum_{k=1}^{S} \sigma_k \qquad (1)$$

where $\sigma_k = \pm 1$ is the random step variable (increment), assigned by the following probabilistic update rule,

$$P(\sigma_{S+1} = \pm 1 | \Delta_S) = \frac{1}{2}\left[1 \pm \left(\frac{\Delta_S}{N+S}\right)\right] \qquad (2)$$

with $N > 0$ (this parameter need not be an integer). The step probabilities depend on the current position (and hence the entire previous step history) as well as the number of steps to date. This construct shares important features with several well-established models which feature long-term memory [9-15]. The broader motivation behind these models is to study mechanisms which generate anomalous diffusive behaviour at large scale, as found widely in the physical and life sciences [16, 17]. In that context, the issue of record statistics is equally as interesting as for processes which are characterized by conventional diffusive behaviour; see also the discussion at the end of the paper. The limit $N \to 0$ in (2) is trivial, *in extremis*, since every step has the same sign as the first step (the motion is ballistic), but the behaviour when $N \neq 0$ is far from trivial. The limit $N \to \infty$ corresponds to the simple random walk (SRW) with independent and identically distributed (i.i.d.) steps, in a sense to be made precise by various examples below.

In what follows the focus will be on upper records; these are registered when $\Delta_S$ reaches a new maximum value for the first time (see figure 1). By convention, the first record occurs at time zero. Let us suppose for a given walk there are $M$ records over an observation window $[0,T]$ such that $M \in \{1,2,...,T+1\}$. If one defines $\Delta_{max} \equiv \max\{\Delta_S; 0 \leq S \leq T\}$, simple counting arguments show that $M = \Delta_{max} + 1$ [8]. The central quantity of interest is the probability $P_T(M) \equiv P(\Delta_{max} = M-1)$ of there being precisely $M$ records. The behaviour of this quantity, which is computed explicitly for this model, is very different to the corresponding quantity for the SRW.

For example, the transient nature of the walk means that as $T \to \infty$ the number of records may not grow indefinitely; the formal statement here is $\Sigma_{M=1}^{\infty} P_{\infty}(M) = \frac{1}{2}$ where $P_{\infty}(M) \equiv \lim_{T \to \infty} P_T(M)$ for fixed $M$. The implication is that records may *never* be broken however long one observes for. Moreover, one can obtain a precise result for $P_{\infty}(M)$, which is non-zero. The corresponding result for the SRW (or indeed any recurrent walk) is strictly zero.

The paper is structured as follows. In Section 2, the key characteristics of the model are summarised. In Section 3, exact results are derived in relation to the number of records for any given observation window, together with the asymptotic behaviour over a long observation window. In Section 4, certain results are presented in relation to the duration of records. Finally, in Section 5, conclusions are drawn and suggestions for future directions of enquiry are given.

**2. Key characteristics of the model**

Let us consider the evolution of walks under the action of (1, 2) with starting point $\Delta_0 = 0$. It will prove useful to consider walks conditioned on an intermediary value, and $\langle \Delta_S | \Delta_R \rangle$ and $\langle \Delta_S^2 | \Delta_R \rangle$ denote conditional expectations over all realizations for a given *fixed* value $\Delta_R$, for $S \geq R$. For the former one has from (1, 2) the elementary difference equation,

$$\langle \Delta_{S+1} | \Delta_R \rangle = \langle (\Delta_S + \sigma_{S+1}) | \Delta_R \rangle = \left[ 1 + \frac{1}{N+S} \right] \langle \Delta_S | \Delta_R \rangle.$$

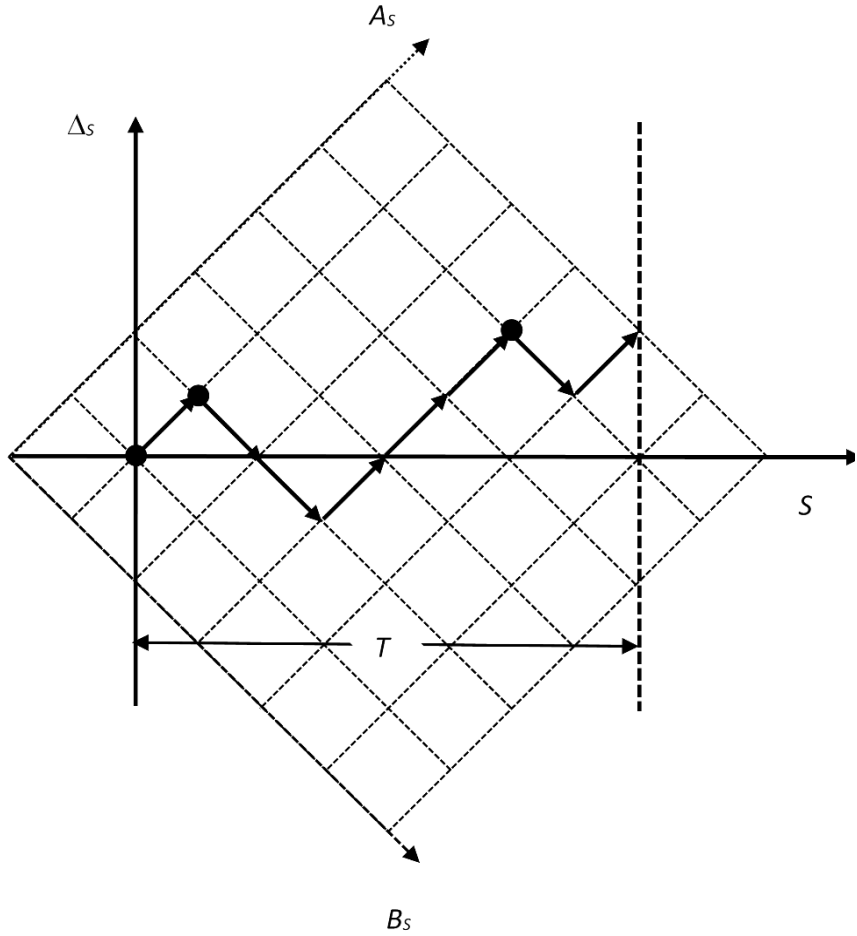

**Figure 1.** A realization of the random walk process $\Delta_S$, with $N=2$, over an observation window of $T=8$ steps. Here, $\Delta_{max}=2$ and the solid dots denote the points at which a new record is set. In the $(A_S, B_S)$ reference frame, the directed path starts at $(1,1)$ and ends at $(6,4)$.

To solve one can simply iterate to derive,

$$\langle \Delta_S | \Delta_R \rangle = \left( \frac{N+S}{N+R} \right) \Delta_R. \tag{3}$$

By setting $R = 0$ one obtains $\langle \Delta_S \rangle = 0$, a result which follows on grounds of symmetry. For the expectation of the square of the displacement one has,

$$\langle \Delta_{S+1}^2 | \Delta_R \rangle = \langle (\Delta_S + \sigma_{S+1})^2 | \Delta_R \rangle = \left[1 + \frac{2}{N+S}\right]\langle \Delta_S^2 | \Delta_R \rangle + 1$$

noting that $\sigma_S^2 \equiv 1$. This is a slightly more complicated difference equation but there are well established techniques for finding the solution (details of which may be found, for example, in the appendix of [15]), with the result that,

$$\langle \Delta_S^2 | \Delta_R \rangle = \frac{(S-R)(N+S)}{N+R+1} + \left(\frac{N+S+1}{N+R+1}\right)\left(\frac{N+S}{N+R}\right)\Delta_R^2. \tag{4}$$

By setting $R = 0$ one obtains;

$$\langle \Delta_S^2 \rangle = \frac{S(N+S)}{N+1}. \tag{5}$$

Since $\langle \Delta_S^2 \rangle = O(S^2)$ as $S \to \infty$ the behaviour is strongly super-diffusive. Note that the limiting behaviour is dependent on $N$.

Concerning the strength of the correlations between steps, one can multiply (3) through by $\Delta_R$ and then average over all realizations using (5);

$$\langle \Delta_S \Delta_R \rangle = \left( \frac{N+S}{N+R} \right) \langle \Delta_R^2 \rangle = \frac{(N+S)R}{N+1}. \qquad (6)$$

This correlator is a product of two functions, one evaluated at time $S$ and the other at time $R$, rather than a single function of the time difference $S-R$. However, one also has from (5, 6) that,

$$\langle (\Delta_S - \Delta_R)^2 \rangle = \frac{(N+S-R)(S-R)}{N+1} = \langle \Delta_{S-R}^2 \rangle$$

which implies there is an underlying stationarity of sorts. It further follows using (6) that for $S > R$,

$$\langle \sigma_S \sigma_R \rangle \equiv \langle (\Delta_S - \Delta_{S-1})(\Delta_R - \Delta_{R-1}) \rangle = \frac{1}{N+1}. \qquad (7)$$

It is a feature of this model that the step correlations do not decay, i.e. they remain persistent for all times.

To characterize the asymptotic behaviour as $S \to \infty$, given that (5) implies that typically $|\Delta_S| = O(S)$, it is helpful to consider the formal sequence of functions,

$$\zeta_n(s) \equiv \frac{1}{n} \sum_{k=1}^{\lfloor ns \rfloor} \sigma_k = \frac{1}{n} \Delta_{\lfloor ns \rfloor}; \qquad \lim_{n \to \infty} \zeta_n(s) \to \zeta(s) \qquad s \in [0,1]$$

which is associated with a given walk. Based on (3) and (4) one has for $s \geq r > 0$;

$$\langle \zeta(s) | \zeta(r) \rangle = \left(\frac{s}{r}\right) \zeta(r); \qquad \langle \zeta(s) \rangle = 0 \qquad (8)$$

$$\langle \zeta(s)^2 | \zeta(r) \rangle = \left(\frac{s}{r}\right)^2 \zeta(r)^2; \qquad \langle \zeta(s)^2 \rangle = \frac{s^2}{N+1}. \qquad (9)$$

From these results it follows that the conditional variance is zero;

$$\mathrm{Var}\left[\zeta(s) | \zeta(r)\right] \equiv \langle \zeta(s)^2 | \zeta(r) \rangle - \langle \zeta(s) | \zeta(r) \rangle^2 = 0.$$

The precise interpretation of this striking result is that,

$$\zeta(s) = Ys; \qquad \Rightarrow \qquad \lim_{S \to \infty} \frac{\Delta_S}{S} = Y \qquad (10)$$

where $Y \in [-1,1]$ is a *realization specific*, non-degenerate random variable whose value for a given walk is principally determined by its initial steps. As will be demonstrated shortly, the probability density of $Y$ is given by,

$$p_Y(y,N) = \frac{\Gamma(N)}{2^{N-1}\Gamma(\tfrac{1}{2}N)\Gamma(\tfrac{1}{2}N)} (1-y^2)^{\tfrac{1}{2}N-1}. \qquad (11)$$

This is, up to a simple transformation, the beta distribution (and specifically *not* a Gaussian). There is a transition in the density profile as $N$ varies from being 'concave' ($N > 2$) to 'convex' ($N < 2$) at the value $N = 2$ (where the density is uniform).[1] The first two moments are given by,

$$\langle Y \rangle = 0; \qquad \langle Y^2 \rangle = \frac{1}{N+1}.$$

As discussed earlier, the limit $N \to 0$ corresponds to ballistic motion with limiting density $p_Y(y, N \to 0) \to \frac{1}{2}\delta(y+1) + \frac{1}{2}\delta(y-1)$.

To derive (11) one can make use of a feature of this type of walk, namely the fact that every possible walk between a given start point and end point has the *same* probabilistic weight. If one considers the set of walks which start at $\Delta_0 = 0$ and finish after $S$ steps at a particular end point $\Delta_S = 2i - S$, where $i$ is chosen from the possible set of values $i = 0, 1, ..., S$, the probabilistic weight of any such realisation of the walk under the action of (2) is given by,

$$W(i, S | \Delta_0 = 0) = \frac{\Gamma(N)}{\Gamma(\frac{1}{2}N)\Gamma(\frac{1}{2}N)} \frac{\Gamma(\frac{1}{2}N + i)\Gamma(\frac{1}{2}N + S - i)}{\Gamma(N + S)}. \tag{12}$$

The end-point probability or propagator $P(i, S | \Delta_0 = 0)$ is then simply given by,

$$P(i, S | \Delta_0 = 0) = \frac{S!}{i!(S-i)!} \times W(i, S | \Delta_0 = 0). \tag{13}$$

---

[1] The explicit dependence on $N$ is a legacy of the fact that the first few steps are strongly determining and memory about them persists indefinitely.

The density (11) follows from (13) in the limit $S \to \infty$. An insightful way to prove (12) is to introduce the variables $A_S$ and $B_S$ (which need not be integers) such that,

$$A_S = \tfrac{1}{2}(N + S + \Delta_S); \qquad B_S = \tfrac{1}{2}(N + S - \Delta_S).$$

Clearly one has $A_S + B_S = N + S$ and $A_S - B_S = \Delta_S$, so with each step *either* $A_S$ or $B_S$ increases by one (figure 1 has a graphical representation for the case where $N = 2$). Then (1, 2) may be rewritten as,

$$P[(A_S, B_S) \to (A_S + 1, B_S)] = \frac{A_S}{A_S + B_S} = \frac{A_S}{A_0 + B_0 + S}$$

$$P[(A_S, B_S) \to (A_S, B_S + 1)] = \frac{B_S}{A_S + B_S} = \frac{B_S}{A_0 + B_0 + S}.$$
(14)

This has the structure of a Pólya urn model with $A_0 = B_0 = \tfrac{1}{2} N$ [18]. The evolution of $A_S$ and $B_S$, taken as a pair, defines a directed path in the $(A_S, B_S)$ reference frame (see figure 1). It is well-known that every path between two given urn states under the action of (14) has the same weight, see e.g. [19],

$$W[(A_0, B_0) \to (A_S, B_S)] = \frac{\Gamma(A_0 + B_0)}{\Gamma(A_0)\Gamma(B_0)} \times \frac{\Gamma(A_S)\Gamma(B_S)}{\Gamma(A_S + B_S)}. \tag{15}$$

After a change of variables, (12) follows. Incidentally, from (14) one can construct an alternative proof of (10) based on martingale convergence [18].

It was pointed out earlier that the limit $N \to \infty$ corresponds to the SRW, although care is needed when the limit $S \to \infty$ is also implied. Thus as $N \to \infty$ one has from (5) that $\langle \Delta_S^2 \rangle = S$, which is the well-known diffusive result. Likewise, (3, 4) reduce to $\langle \Delta_S | \Delta_R \rangle = \Delta_R$ and $\langle \Delta_S^2 | \Delta_R \rangle = S - R + \Delta_R^2$, and (7) reduces to $\langle \sigma_S \sigma_R \rangle = 0$, all as expected. Further, (12) reduces to $W(i, S | \Delta_0 = 0) = (\tfrac{1}{2})^S$, again as expected. More subtly, (11) reduces to $p_Y(y, N \to \infty) \to \delta(y)$. This is an embodiment of the law of large numbers for i.i.d. variables, i.e. $\lim_{S \to \infty} \Delta_S / S \to 0$.

## 3. The distribution and moments of the number of records

The goal is to determine the distribution and moments of the number of records $M$ over an observation window $[0, T]$. The presence of correlated steps means that certain mathematical tools used to analyse the independent step case, such as the Sparre Anderson theorem [2, 20, 21] and ideas from the theory of renewal processes [5, 8], are no longer readily applicable. However, by exploiting ideas discussed in the previous section one can obtain exact results.

To calculate the probability $P_T(M)$ of precisely $M$ records up to $T$ steps, the starting point is the probability $P(\Delta_{\max} > Q)$ for $Q = 0, 1, 2, \ldots$, in terms of which,

$$P_T(M) = P(\Delta_{\max} > M - 2) - P(\Delta_{\max} > M - 1). \tag{16}$$

Since the weight of a given walk (12) depends only on its end point, one can write,

$$P(\Delta_{max} > Q) = \sum_{i=0}^{T} N^{\#}(Q,T,i) \times W(i,T|\Delta_0 = 0)$$

where $N^{\#}(Q,T,i)$ is the number of distinct walks which start at $\Delta_0 = 0$ and end at $\Delta_T = 2i - T$ for which $\Delta_{max} > Q$. If $\Delta_T > Q$ then clearly all walks must satisfy $\Delta_{max} > Q$. If $\Delta_T \leq Q$, then only a subset of walks will satisfy $\Delta_{max} > Q$, the combinatorial evaluation of which is aided by making use of the reflection principle. After some relatively routine algebra one has,

$$N^{\#}(Q,T,i) = \begin{cases} \dfrac{T!}{i!(T-i)!}; & i \geq \left\lfloor \dfrac{T+Q}{2} \right\rfloor + 1 \\[2ex] \dfrac{T!}{(i-Q-1)!(T-i+Q+1)!}; & i \leq \left\lfloor \dfrac{T+Q}{2} \right\rfloor. \end{cases}$$

It follows in turn that,

$$P(\Delta_{max} > Q) = \frac{\Gamma(N)}{\Gamma(\tfrac{1}{2}N)\Gamma(\tfrac{1}{2}N)} \left[ F_+(Q,T) + F_-(Q,T) \right] \tag{17}$$

where,

$$F_+(Q,T) = \frac{T!}{\Gamma(T+N)} \sum_{i=\left\lfloor \frac{T+Q}{2} \right\rfloor + 1}^{T} \frac{\Gamma(\tfrac{1}{2}N+i)\Gamma(\tfrac{1}{2}N+T-i)}{i!(T-i)!}$$

$$F_-(Q,T) = \frac{T!}{\Gamma(T+N)} \sum_{i=Q+1}^{\left\lfloor \frac{T+Q}{2} \right\rfloor} \frac{\Gamma(\tfrac{1}{2}N+i)\Gamma(\tfrac{1}{2}N+T-i)}{(i-Q-1)!(T-i+Q+1)!}.$$

(18)

One can show that $P(\Delta_{max} > -1) \equiv 1$ using the Chu-Vandermonde identity [22], as must be the case since $\Delta_0 = 0$. In conjunction with (16) one can now calculate $P_T(M)$ and its associated moments. For the first moment of $M$ one has;

$$\langle M \rangle = \sum_{M=1}^{T+1} M \times P_T(M) = 1 + \sum_{Q=0}^{T-1} P(\Delta_{max} > Q). \qquad (19)$$

In the limit $N \to 0$ this simplifies to $\langle M \rangle = 1 + \tfrac{1}{2}T$, which comes from there being only two possible outcomes, $M = T+1$ or $M = 1$, depending on the initial random step, and each outcome occurs with probability $\tfrac{1}{2}$. As an illustration, results for the first few values of $T$ for the case $N = 2$, which is the concave/convex transition point in the density profile (11), are given in Table 1.

The corresponding values for the SRW are also given in Table 1. One sees that, for $T > 1$, the first moment is enhanced due to the positive correlations between steps. To compute the SRW results one can first take the limit $N \to \infty$ in (17, 18) with $T$ fixed to derive from (16) the concise result,

$$P_T(M) = \frac{1}{2^T} \frac{T!}{\left(\left\lceil \frac{T+M-1}{2} \right\rceil\right)! \left(\left\lfloor \frac{T-M+1}{2} \right\rfloor\right)!} \qquad (20)$$

which is known by other means, see e.g. [8].

| $\langle M \rangle$ | $T=0$ | $T=1$ | $T=2$ | $T=3$ | $T=4$ | $T=5$ | $T=6$ |
|---|---|---|---|---|---|---|---|
| $N=0$ | 1 | $\frac{3}{2}$ | 2 | $\frac{5}{2}$ | 3 | $\frac{7}{2}$ | 4 |
| $N=2$ | 1 | $\frac{3}{2}$ | $\frac{11}{6}$ | $\frac{13}{6}$ | $\frac{37}{15}$ | $\frac{83}{30}$ | $\frac{641}{210}$ |
| SRW ($N=\infty$) | 1 | $\frac{3}{2}$ | $\frac{7}{4}$ | 2 | $\frac{35}{16}$ | $\frac{19}{8}$ | $\frac{81}{32}$ |

**Table 1**: The first moment of the number of records $M$.

Concerning the limiting behaviour as $T \to \infty$, the dominant contribution to the moments for $N$ fixed comes from the term involving the function $F_+(Q,T)$ in (18). Regarding the first moment, one may show using (17) and (19) that,

$$\langle M \rangle_+ \equiv \frac{\Gamma(N)}{\Gamma(\frac{1}{2}N)\Gamma(\frac{1}{2}N)} \sum_{Q=0}^{T-1} F_+(Q,T)$$

$$\approx \frac{\Gamma(N)}{\Gamma(\frac{1}{2}N)\Gamma(\frac{1}{2}N)} \int_0^T \int_{\frac{1}{2}+\frac{Q}{2T}}^1 x^{\frac{1}{2}N-1}(1-x)^{\frac{1}{2}N-1} dx dQ \qquad (21)$$

$$= \frac{\Gamma(N+1)}{2^{N+1}\Gamma(\frac{1}{2}N+1)\Gamma(\frac{1}{2}N+1)} T + O(1).$$

In this derivation, use is made of the Euler-Maclaurin formula and the fact that $\Gamma(z+c)/\Gamma(z+d) \sim z^{c-d}$ as $z \to \infty$. The growth is linear in $T$, which loosely mirrors what happens in a random walk with independent steps but with positive (persistent) overall drift [23].

At a fundamental level, this term is associated with walks for which $\Delta_S \to +\infty$, and there is a more powerful embodiment of this observation. It follows from (10) that beyond a certain number of steps (which is realization specific) the displacement $\Delta_S$ is either permanently positive or negative (transient behaviour). Now if $\Delta_S \to +\infty$ ($Y > 0$) the number of records will grow unbounded as $S \to \infty$. More precisely, over an observation widow $[0,T]$, one has that $P(\lim_{T\to\infty}(\Delta_{\max} - \Delta_T)/T > \varepsilon) = 0$ for all $\varepsilon > 0$, i.e. $\Delta_{\max}/T \to Y$. Recalling that $M = \Delta_{\max} + 1$, this implies that $M/T \to Y$. Conversely, if $\Delta_S \to -\infty$ ($Y < 0$) the number of records will saturate to a finite value, which implies that $M/T \to 0$. As a result,

$$\lim_{T \to \infty} \frac{\langle M^k \rangle}{T^k} = \gamma_k; \qquad \gamma_k \equiv \int_0^1 y^k p_Y(y,N)\, dy. \tag{22}$$

Thus, using (11), one can derive the asymptotic behaviour of all the moments, e.g.,

$$\langle M \rangle \sim \frac{\Gamma(N+1)}{2^{N+1}\Gamma(\tfrac{1}{2}N+1)\Gamma(\tfrac{1}{2}N+1)} T; \qquad \langle M^2 \rangle \sim \frac{T^2}{2(N+1)}. \tag{23}$$

As expected, given the strong step correlations, these are markedly different from the corresponding results for the SRW which follow from (19, 20);

$$\langle M \rangle \sim \sqrt{\frac{2T}{\pi}}; \qquad \langle M^2 \rangle \sim T.$$

One cannot recover these results by taking the limit $N \to \infty$ in (23) because the limit $T \to \infty$ has been taken first; in this instance the limits are not interchangeable. One can, however, derive scaling forms which effectively interpolate by analysing (16-19) from the perspective of letting $T, N \to \infty$ with the ratio $T/N$ held *fixed*. After a straightforward but lengthy calculation one finds, for example,

$$\langle M \rangle \sim \sqrt{N} f_1\left(\frac{T}{N}\right); \qquad f_1(x) = \frac{1}{\sqrt{2\pi}}\left[\sqrt{x(1+x)} + \log\left(\sqrt{1+x} + \sqrt{x}\right)\right].$$

The logarithmic term in $f_1(x)$ arises from the term $F_-(Q,T)$ in (18), which is interesting and is discussed further below. It is easy to check that the relevant scaling limits are recovered in the cases where $T/N \gg 1$ and $T/N \ll 1$.

Although the term $F_-(Q,T)$ does not contribute to the leading order asymptotic behaviour of the moments as $T \to \infty$ with $N$ fixed, it does have an important role to play in other regards, as evidenced above. This term is associated with walks for which $\Delta_S \to -\infty$. After a more challenging exercise than the derivation of (21), since greater care is now needed to treat the twin limits $T \to \infty$ and $Q \to \infty$ correctly, one can show that the contribution of $F_-(Q,T)$ to the first moment for $N$ fixed (after making use of symmetry to halve the domain of the double summation) is,

$$\langle M \rangle_{-} \equiv \frac{\Gamma(N)}{\Gamma(\frac{1}{2}N)\Gamma(\frac{1}{2}N)} \sum_{Q=0}^{T-1} F_{-}(Q,T)$$

$$\approx 2 \times \frac{\Gamma(N)}{\Gamma(\frac{1}{2}N)\Gamma(\frac{1}{2}N)} \int_{0}^{\frac{T}{2}} \int_{0}^{\frac{1}{2}-\frac{Q}{T}} x^{\frac{1}{2}N+Q}(1-x)^{\frac{1}{2}N-Q-2} \, dx \, dQ \qquad (24)$$

$$= \frac{\Gamma(N)}{2^{N}\Gamma(\frac{1}{2}N)\Gamma(\frac{1}{2}N)} \log T + O(1).$$

Combined with (21) one therefore has,

$$\langle M \rangle = \frac{\Gamma(N+1)}{2^{N+1}\Gamma(\frac{1}{2}N+1)\Gamma(\frac{1}{2}N+1)} T + \frac{\Gamma(N)}{2^{N}\Gamma(\frac{1}{2}N)\Gamma(\frac{1}{2}N)} \log T + O(1).$$

The logarithmic term is related to the fact that $\lim_{T \to \infty} P_T(M) \neq 0$ for $M$ fixed. This is a fundamental consequence of the transience of the process and is quite different to the case of the SRW where, due to the recurrence of the process, any fixed upper value will eventually be exceeded given enough time and so $\lim_{T \to \infty} P_T(M) = 0$, as may be seen from (20). To explore this special feature further, as $T \to \infty$ one can evaluate the functions $F_{+}(Q,T)$ and $F_{-}(Q,T)$ in (18) thus;

$$F_{+}(Q, T \to \infty) = \int_{\frac{1}{2}}^{1} x^{\frac{1}{2}N-1}(1-x)^{\frac{1}{2}N-1} dx = \frac{\Gamma(\frac{1}{2}N)\Gamma(\frac{1}{2}N)}{2\Gamma(N)}$$

$$F_{-}(Q, T \to \infty) = \int_{0}^{\frac{1}{2}} x^{\frac{1}{2}N+Q}(1-x)^{\frac{1}{2}N-Q-2} dx = \int_{0}^{1} \frac{z^{\frac{1}{2}N+Q}}{(1+z)^{N}} dz.$$

These results are exact, despite the impression that approximations have been made (the error terms vanish in the limit). It follows in conjunction with (17) that,

$$\lim_{T\to\infty} P(\Delta_{max} > Q) = \frac{1}{2} + \frac{\Gamma(N)}{\Gamma(\frac{1}{2}N)\Gamma(\frac{1}{2}N)} \int_0^1 \frac{z^{\frac{1}{2}N+Q}}{(1+z)^N} dz. \quad (25)$$

In this way, using (16), one can define $P_\infty(M) \equiv \lim_{T\to\infty} P_T(M)$ so that,

$$P_\infty(M) = \frac{\Gamma(N)}{\Gamma(\frac{1}{2}N)\Gamma(\frac{1}{2}N)} \int_0^1 \frac{z^{\frac{1}{2}N+M-2}(1-z)}{(1+z)^N} dz. \quad (26)$$

As discussed in the Introduction, evidently this is non-zero. One can further note based on (16) and (25) that $\Sigma_{M=1}^\infty P_\infty(M) = \frac{1}{2}$, which encapsulates the transience of the process and the symmetry of the initial condition. It is intuitively clear on physical grounds that $P_\infty(M)$ must decrease to zero as $M \to \infty$ and from (26) one has,

$$P_\infty(M) \sim \frac{\Gamma(N)}{2^N \Gamma(\frac{1}{2}N)\Gamma(\frac{1}{2}N)} \frac{1}{M^2}.$$

This limiting behaviour underpins the origin of the $\log T$ term in (24).

One can derive alternative expressions to (25, 26) by integrating by parts indefinitely;

$$\lim_{T \to \infty} P(\Delta_{max} > Q) = \frac{1}{2} + \frac{\Gamma(\frac{1}{2}N + Q + 1)}{2^{N-1}\Gamma(\frac{1}{2}N)\Gamma(\frac{1}{2}N)} \sum_{j=1}^{\infty} \frac{\Gamma(N-1+j)}{2^{j}\Gamma(\frac{1}{2}N + Q + j + 1)}$$

$$P_{\infty}(M) = \frac{\Gamma(\frac{1}{2}N + M - 1)}{2^{N-1}\Gamma(\frac{1}{2}N)\Gamma(\frac{1}{2}N)} \sum_{j=1}^{\infty} \frac{j\Gamma(N-1+j)}{2^{j}\Gamma(\frac{1}{2}N + M + j)}.$$

By integrating by parts a different way, it is also possible to obtain a recursion which can be solved in terms of a finite summation. In general, the resulting expression is complicated and unwieldy but by way of illustration one has for the case $N = 2$ (where a significant simplification occurs);

$$\lim_{T \to \infty} P(\Delta_{max} > Q) = \frac{1}{2} + (-1)^{Q+1}(Q+1)\left[\frac{1}{2} + \frac{1}{2}\sum_{j=1}^{Q} \frac{(-1)^{j-1}}{j(j+1)} - \log 2\right]$$

$$P_{\infty}(M) = (-1)^{M-1}(2M-1)\left[\frac{1}{2} + \frac{1}{2}\sum_{j=1}^{M-1} \frac{(-1)^{j-1}}{j(j+1)} - \log 2\right] + \frac{1}{2M}$$

from whence it follows (for $N = 2$) that,

$$P_{\infty}(M = 1) = 1 - \log 2 = 0.306...$$

$$P_{\infty}(M = 2) = 3\log 2 - 2 = 0.079...$$

$$P_{\infty}(M = 3) = \tfrac{7}{2} - 5\log 2 = 0.034...$$

and so on.

## 4. The duration of records

For the case $M = 1$ one has from evaluating (26),

$$P_\infty(M = 1) = \left(\frac{1}{N-2}\right)\left[\frac{\Gamma(N)}{\Gamma(\tfrac{1}{2}N)\Gamma(\tfrac{1}{2}N)}\left(\frac{1}{2}\right)^{N-2} - 1\right]; \quad N \neq 2$$

$$= 1 - \log 2; \quad N = 2. \tag{27}$$

As noted above, this quantity, which is the probability that there is only one record however long one observes, is non-zero. The implication is that a given walk, starting at zero and moving downwards at the first step, may never actually cross the line $\Delta_S = 0$ from below (although it may return to it without crossing).

The statement that there is only one record is the same as saying the first record has unlimited duration. To explore this further, let us evaluate the duration $l_1$ of the first record given unlimited observation time. Consider a walk which starts at $\Delta_0 = 0$ and after $l_1 = 2n+1$ steps reaches $\Delta_S = 1$ for the first time (first passage through zero), where $n = 0, 1, 2, \ldots$. In the $(A_S, B_S)$ reference frame the directed path starts at $(\tfrac{1}{2}N, \tfrac{1}{2}N)$ and ends at $(\tfrac{1}{2}N + n + 1, \tfrac{1}{2}N + n)$, and is constrained to satisfy $A_S \leq B_S$ at all points in-between. Adapting what was discussed in the previous section, the number of such paths may be derived using the reflection principle and is simply the Catalan number $C_n$. Since every contributing path has the same weight it follows that,

$$P(l_1 = 2n+1) = C_n \times W\left[(\tfrac{1}{2}N, \tfrac{1}{2}N) \to (\tfrac{1}{2}N+n+1, \tfrac{1}{2}N+n)\right]$$

where the weight function is given by (15). The probability that the first record has a duration no greater than $2L+1$ is thus given by,

$$P(l_1 \leq 2L+1) = \frac{\Gamma(N)}{\Gamma(\tfrac{1}{2}N)\Gamma(\tfrac{1}{2}N)} \sum_{j=0}^{L} \frac{(2j)!\Gamma(\tfrac{1}{2}N+j+1)\Gamma(\tfrac{1}{2}N+j)}{(j+1)!j!\Gamma(N+2j+1)}. \quad (28)$$

The probability that the duration is *finite* is $P(l_1 < \infty) \equiv \lim_{L \to \infty} P(l_1 \leq 2L+1)$, which numerical evaluation shows is less than unity. This is the signature of a defective process, meaning the probability of unlimited duration is non-zero, and by definition $P(l_1 = \infty) \equiv 1 - P(l_1 < \infty)$.[2] Naturally one must have $P(l_1 = \infty) = P_\infty(M=1)$, where the latter quantity is given by (27). Comparing (27) and $P(l_1 = \infty)$ as derived from (28) gives a non-trivial identity which can be confirmed numerically for general $N$. For the case $N=2$ one has a direct demonstration; thus,

$$P(l_1 \leq 2L+1) = \sum_{j=0}^{L} \frac{1}{(2j+1)(2j+2)} = \sum_{j=0}^{2L+1} \frac{(-1)^j}{j+1}$$

$$P(l_1 < \infty) \equiv \lim_{L \to \infty} P(l_1 \leq 2L+1) = \log 2$$

$$P(l_1 = \infty) = 1 - \log 2.$$

---

[2] It should be stressed that $P(l_1 = \infty) \neq \lim_{n \to \infty} P(l_1 = 2n+1) \to 0$.

A straightforward extension of this line of reasoning shows the probability that the duration $l_r$ of the $r$-th record is unlimited (given the preceding records are of finite duration) is given by $P(l_r = \infty) = P_\infty(M = r)$, where the latter quantity is given by (26). One can push the analysis to make more refined statements about the duration of subsequent records, but the effort involved is significant for no real new insight gained and so this is not pursued any further here.

In the limit $N \to \infty$ one can again recover the relevant results for the SRW. Thus, for fixed $L$ (28) reduces as $N \to \infty$ to,

$$P(l_1 \leq 2L+1) = \sum_{j=0}^{L} \left(\frac{1}{2}\right)^{2j+1} \frac{(2j)!}{(j+1)!\,j!}.$$

Using a well-known identity for the Catalan numbers one then has $P(l_1 < \infty) = 1$ and the process is not defective. In other words, for the SRW all records have finite duration with probability one, which reflects the recurrence of the process.

**5. Conclusions**

In this paper, a random walk model with correlated steps has been studied in relation to record statistics over a given observation window, with various exact results presented, notably (22) and (26). Although relatively simple in construct, the model has interesting behaviour and illustrates the key message that, when the step correlations are sufficiently strong, the record statistics are distinctly different from those for a random walk with independent steps.

One can imagine extending this work by generalising the probabilistic update rule (2) to align more closely with the models discussed in [9-15], all of which exhibit a transition (as an additional parameter is varied) between a regime where the behaviour is diffusive and a regime where the behaviour is super-diffusive. As an illustration, following [15] one such modification of (2) is;

$$\Pr(\sigma_{S+1} = \pm 1 | \Delta_S) = \frac{1}{2}\left[1 \pm \alpha\left(\frac{\Delta_S}{N+S}\right)\right]$$

with $0 \leq \alpha \leq 1$. The case $\alpha = 1$ corresponds to (2), the case $\alpha = 0$ corresponds to the SRW from a different perspective than the limit $N \to \infty$, whilst the limit $N \to 0$ corresponds to the so-called elephant random walk model with global memory parameter $p = \frac{1}{2}(1+\alpha)$ and initial step parameter $q = \frac{1}{2}$ [9]. As before $\langle \Delta_S \rangle = 0$ but now (5) becomes,

$$\langle \Delta_S^2 \rangle = \frac{N+S}{2\alpha - 1}\left[\frac{\Gamma(N+1)}{\Gamma(N+2\alpha)}\frac{\Gamma(N+2\alpha+S)}{\Gamma(N+1+S)} - 1\right].$$

In the limit $S \to \infty$ one has $\langle \Delta_S^2 \rangle = O(S)$ when $\alpha < \frac{1}{2}$, but $\langle \Delta_S^2 \rangle = O(S^{2\alpha})$ when $\alpha > \frac{1}{2}$. Thus, the limiting dynamics are diffusive when $\alpha < \frac{1}{2}$ (with Hurst exponent $H = \frac{1}{2}$) but super-diffusive when $\alpha > \frac{1}{2}$ (with $H = \alpha$). By such means, one might study the transition in record statistics as the strength of the step correlations changes from being weak to being strong. In such models, however, the weight of a given walk is typically rendered fully path dependent, whereupon much of the analysis presented earlier breaks down. An exact calculation of the record statistics over a finite observation window $[0,T]$ is likely to be a much more challenging task in such circumstances. Even to gain insight into the asymptotic behaviour as $T \to \infty$ may require fresh ideas and techniques.